\documentclass[10pt,twocolumn]{article}

\usepackage[letterpaper,top=0.62in,bottom=0.68in,left=0.70in,right=0.70in,columnsep=0.22in]{geometry}
\usepackage{amsmath,amssymb}
\usepackage{graphicx}
\usepackage{array}
\usepackage{booktabs}
\usepackage{microtype}
\usepackage{caption}
\usepackage{fancyhdr}
\usepackage[hidelinks]{hyperref}

\newcommand{\emailmark}[1]{\textsuperscript{#1}}
\newcommand{\runin}[1]{\par\noindent\textit{#1}---\ }

\begin{document}

\twocolumn[
\begin{@twocolumnfalse}
\begin{center}
{\Large\bfseries Supernova Time Dilation in Hybrid Expansion--Tired-Light Cosmologies\par}
\vspace{7pt}
{\normalsize Rajendra P. Gupta\textsuperscript{1,*}, Nikolaos Samaras\textsuperscript{1,\ensuremath{\dagger}}, and Utkarsh Kumar\textsuperscript{1,\ensuremath{\ddagger}}\par}
\vspace{3pt}
{\small\itshape\textsuperscript{1}Department of Physics, University of Ottawa, Ottawa, Ontario K1N 6N5, Canada\par}
\end{center}
\vspace{5pt}
\begin{center}
\begin{minipage}{0.84\textwidth}
\small
Dark Energy Survey has shown that the emission light curve widths $\Delta t_{\rm em}$ of supernovae increase to $\Delta t_{\rm obs}$, obeying
$\Delta t_{\rm obs}/\Delta t_{\rm em}=(1+z)^b$ with
$b=1.003\pm0.011$, excluding a nondilating redshift.
We test hybrid models with $1+z=(1+z_p)(1+z_t)$ and the phenomenological stretch
$(1+z_p)R(z)$, where $z_p$ is the redshift associated with the expanding parent universe cosmology, $z_t$ is a tired-light (TL) contribution, and $R$ is the hybrid-to-parent lookback-time ratio.
At $z=1$, $R$ and $1+z_t$ differ by $1.3\%$ for CCC+TL and $0.5\%$ for $\Lambda$CDM+TL,
and their 95\% bands overlap the observed relation.
This compatibility shows that the observations do not exclude the hybrid models, only fully nonexpanding ones.
\end{minipage}
\end{center}
\vspace{8pt}
\end{@twocolumnfalse}
]
\thispagestyle{empty}

Cosmological time dilation is one of the most direct kinematic signatures of an expanding universe. Type Ia supernova (SN Ia) light curves broaden with redshift, and both photometric and spectroscopic studies have found the expected scaling with $1+z$~\cite{Goldhaber2001,Riess1997,Foley2005,Blondin2008}. Most recently, White et al. analyzed 1504 SNe Ia from the Dark Energy Survey over $0.1\lesssim z\lesssim1.2$ and parametrized the observed duration as~\cite{White2024}
\begin{equation}
\begin{aligned}
\Delta t_{\rm obs}&=\Delta t_{\rm em}(1+z)^b,\\
 b&=1.003\pm0.005\ ({\rm stat})\pm0.010\ ({\rm sys}).
\end{aligned}
\label{eq:des}
\end{equation}
This result rules out, at high significance, models in which the full observed redshift is generated by a mechanism that changes photon energy without stretching arrival-time intervals.

The inference is less immediate for a hybrid model in which only part of the redshift is nondilating. We consider the decomposition introduced in Ref.~\cite{Gupta2023},
\begin{equation}
1+z=(1+z_p)(1+z_t),
\label{eq:split}
\end{equation}
where $z_p$ is the redshift associated with the expanding parent cosmology and $z_t$ is a tired-light (TL) contribution. In the simplest nondispersive TL interpretation, the TL process reduces photon energy but does not by itself broaden a transient. The direct kinematic stretch is then $1+z_p$, leaving an apparent deficit by the factor $1+z_t$ relative to Eq.~(\ref{eq:des}) with $b\simeq1$.

For clarity, the standard expansion result can be obtained by comparing two neighboring radial null rays emitted a proper time $\delta t_{\rm em}$ apart. Equality of their comoving paths gives
\begin{equation}
\frac{\delta t_{\rm obs}}{a(t_0)}
=
\frac{\delta t_{\rm em}}{a(t_e)},
\label{eq:nullrays}
\end{equation}
so that $\delta t_{\rm obs}/\delta t_{\rm em}=a(t_0)/a(t_e)=1+z_p$. This relation depends on the scale-factor ratio, not on the total photon travel time.

\runin{Lookback-time prescription.}
Hybrid models can nevertheless have substantially different age--redshift relations from their parent cosmologies. Let
\begin{equation}
R(z)\equiv
\frac{t_L^{\rm hyb}(z)}{t_L^p(z)},
\label{eq:R}
\end{equation}
where $t_L^{\rm hyb}$ and $t_L^p$ are the lookback times at the same observed redshift in the hybrid and parent models. We examine the phenomenological prescription
\begin{equation}
S_{\rm hyb}(z)\equiv
\frac{\Delta t_{\rm obs}}{\Delta t_{\rm em}}
=(1+z_p)R(z).
\label{eq:stretch}
\end{equation}
Combining Eqs.~(\ref{eq:split}) and (\ref{eq:stretch}), agreement with the $b=1$ limit of Eq.~(\ref{eq:des}) requires
\begin{equation}
R(z)\simeq1+z_t.
\label{eq:condition}
\end{equation}
Equation~(\ref{eq:stretch}) is the additional model prescription tested here. It is not implied by the redshift decomposition alone: a complete theory must derive it, or an alternative interval mapping, from the metric and photon-propagation law. The calculation below therefore tests numerical compatibility conditional on Eq.~(\ref{eq:stretch}).

\runin{Numerical comparison.}
We use the four Pantheon+ fits reported in Ref.~\cite{Gupta2023}: CCC and its hybrid CCC+TL extension, and $\Lambda$CDM and its $\Lambda$CDM+TL extension. The parent model is essential to the comparison. Contrasting CCC+TL directly with $\Lambda$CDM, for example, mixes the effect of adding TL with the differences between the CCC and $\Lambda$CDM backgrounds.

At observed redshift $z=1$, the CCC+TL and CCC lookback times give
\begin{equation}
R^C=1.144\pm0.042,\qquad 1+z_t^C=1.159\pm0.010,
\label{eq:ccc}
\end{equation}
while the $\Lambda$CDM+TL and $\Lambda$CDM comparison gives
\begin{equation}
R^L=1.140\pm0.041,\qquad 1+z_t^L=1.146\pm0.009.
\label{eq:lcdm}
\end{equation}
All quoted intervals are at 95\% confidence. The fractional residual
\begin{equation}
\epsilon(z)\equiv
\frac{S_{\rm hyb}(z)}{1+z}-1
=
\frac{R(z)}{1+z_t}-1
\label{eq:residual}
\end{equation}
is $-1.3\%$ for CCC+TL and $-0.5\%$ for $\Lambda$CDM+TL at $z=1$. The 95\% intervals overlap in both cases. The complete mean-value sequences used in the calculation are given in the Appendix.

\begin{figure}[t]
\centering
\includegraphics[width=\columnwidth]{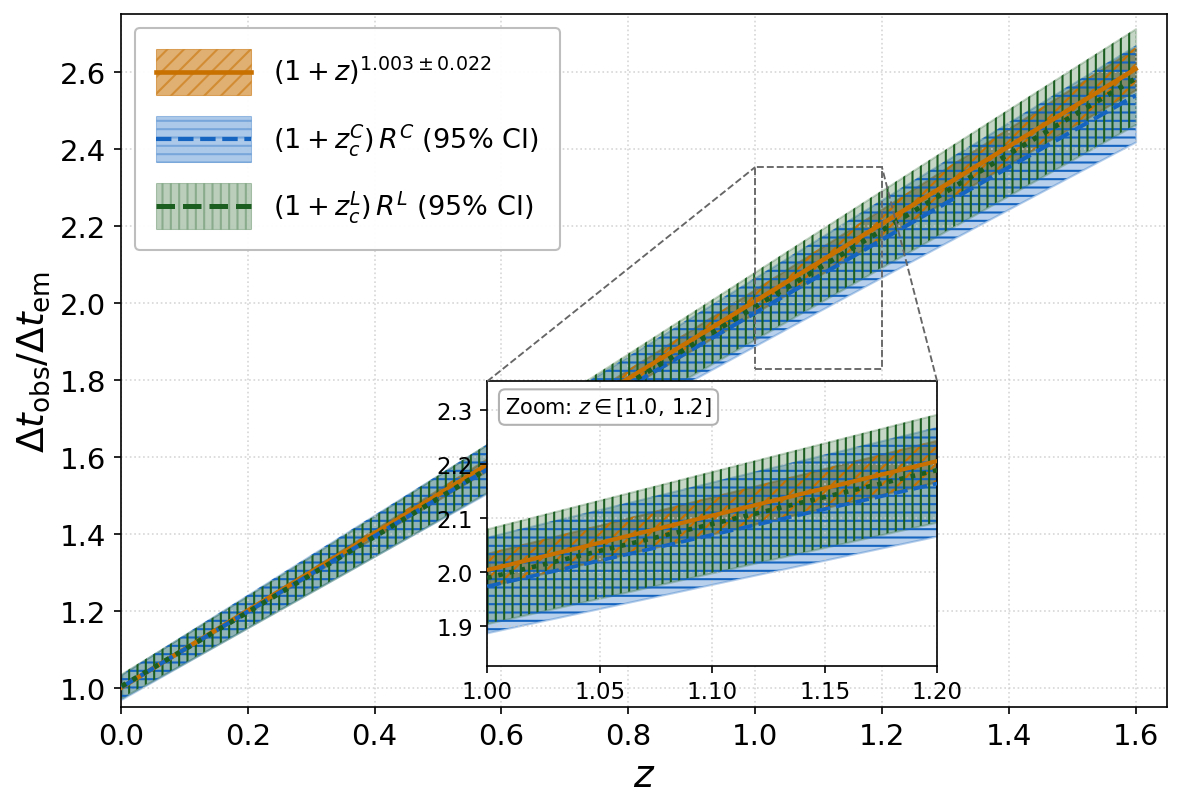}
\caption{Observed time-stretch relation from White et al.~\cite{White2024} (orange) compared with Eq.~(\ref{eq:stretch}) for CCC+TL relative to CCC (blue) and $\Lambda$CDM+TL relative to $\Lambda$CDM (green). Superscripts $C$ and $L$ label the two parent-model comparisons. The bands are shown at 95\% confidence.}
\label{fig:comparison}
\end{figure}

Figure~\ref{fig:comparison} compares the DES relation, with the statistical and systematic uncertainties combined in quadrature and converted to an approximate 95\% interval, with the prediction of Eq.~(\ref{eq:stretch}). The three bands overlap extensively from $z\simeq0$ to 1.6. The two hybrid constructions also track one another closely, indicating that the near-cancellation between $R(z)$ and $1+z_t$ is not unique to the CCC background within these fits.

The result sharpens what SN time-dilation observations do and do not establish. They decisively exclude a fully nondilating interpretation of the observed redshift, because such a model predicts $S=1$. They also constrain a hybrid model if its interval mapping is simply $S=1+z_p$. Under the additional prescription in Eq.~(\ref{eq:stretch}), however, the reduced expansion redshift is nearly compensated by the changed age--redshift relation in both examples considered here. The DES measurement is then numerically consistent with a modest nondilating redshift contribution.

The distinction between numerical compatibility and dynamical explanation is important. The lookback time is an integrated property of the background expansion, whereas the separation of two neighboring photon arrival times is determined locally by the mapping between emission and observation events. A future derivation should begin with the CCC+TL line element and the specified TL transport law, propagate two neighboring wave packets, and obtain $dt_{\rm obs}/dt_{\rm em}$ without inserting $R(z)$ by hand. If that derivation yields Eq.~(\ref{eq:stretch}), the comparison above shows that current SN Ia time-dilation data do not independently rule out the hybrid fits. If it yields only $1+z_p$, the same data provide a strong constraint on the permitted TL fraction.

In summary, the relevant observable for a hybrid redshift model is the complete transient-duration mapping, not the TL factor considered in isolation. For the phenomenological mapping $S_{\rm hyb}=(1+z_p)R(z)$, the CCC+TL and $\Lambda$CDM+TL examples reproduce the DES time-stretch relation to percent-level accuracy over the redshift interval examined. This conditional agreement motivates a first-principles derivation of the interval mapping as the decisive next test.

\smallskip
\noindent We thank Ethan Vishniac for discussions that motivated this work.

\smallskip
\noindent\textit{Data availability.} The numerical values used in this Letter are reproduced in the Appendix and are derived from the Pantheon+ model fits reported in Ref.~\cite{Gupta2023}; the underlying Pantheon+ data are available through Refs.~\cite{Brout2022,Scolnic2022}.

\smallskip
\noindent\emailmark{*}\,\texttt{rgupta4@uottawa.ca}\par
\noindent\emailmark{\ensuremath{\dagger}}\,\texttt{nsamaras@uottawa.ca}\par
\noindent\emailmark{\ensuremath{\ddagger}}\,\texttt{kumarutkarsh641@gmail.com}

\begingroup
\small
\setlength{\itemsep}{0pt}
\setlength{\parsep}{0pt}
\renewcommand{\section}[2]{}

\endgroup

\clearpage
\onecolumn
\begin{center}
{\large\bfseries APPENDIX: NUMERICAL VALUES}
\end{center}
\vspace{-2pt}

The table lists the mean lookback times and TL factors used to construct Fig.~\ref{fig:comparison}. CTL denotes CCC+TL and LTL denotes $\Lambda$CDM+TL. Lookback times are in Gyr.

\begin{table}[h]
\centering
\small
\setlength{\tabcolsep}{8.5pt}
\renewcommand{\arraystretch}{1.08}
\begin{tabular}{c c c c c c c}
\toprule
$z$ & $t_L^{\rm CTL}$ & $t_L^{\rm CCC}$ & $1+z_t^C$ & $t_L^{\rm LTL}$ & $t_L^{\Lambda{\rm CDM}}$ & $1+z_t^L$ \\
\midrule
0.001 & 0.013475 & 0.013461 & 1.000181 & 0.013454 & 0.013406 & 1.000168 \\
0.1   & 1.266935 & 1.245197 & 1.017797 & 1.265538 & 1.245229 & 1.016615 \\
0.2   & 2.391412 & 2.313993 & 1.035100 & 2.388406 & 2.319109 & 1.032763 \\
0.3   & 3.398406 & 3.239873 & 1.051944 & 3.391909 & 3.249199 & 1.048451 \\
0.4   & 4.307102 & 4.048402 & 1.068362 & 4.294251 & 4.058103 & 1.063688 \\
0.5   & 5.132525 & 4.759452 & 1.084381 & 5.109936 & 4.764530 & 1.078483 \\
0.6   & 5.886643 & 5.388715 & 1.100026 & 5.850712 & 5.384037 & 1.092846 \\
0.7   & 6.579130 & 5.948745 & 1.115321 & 6.526233 & 5.929592 & 1.106790 \\
0.8   & 7.217906 & 6.449707 & 1.130285 & 7.144525 & 6.412024 & 1.120327 \\
0.9   & 7.809525 & 6.899916 & 1.144937 & 7.712330 & 6.840390 & 1.133469 \\
1.0   & 8.359470 & 7.306235 & 1.159294 & 8.235365 & 7.222277 & 1.146230 \\
1.1   & 8.872368 & 7.674380 & 1.173370 & 8.718515 & 7.564057 & 1.158623 \\
1.2   & 9.352154 & 8.009143 & 1.187180 & 9.165981 & 7.871096 & 1.170660 \\
1.3   & 9.802200 & 8.314572 & 1.200737 & 9.581400 & 8.147921 & 1.182355 \\
1.4   & 10.225410 & 8.594106 & 1.214053 & 9.967938 & 8.398372 & 1.193722 \\
1.5   & 10.624320 & 8.850686 & 1.227138 & 10.328360 & 8.625709 & 1.204771 \\
1.6   & 11.001130 & 9.086837 & 1.240003 & 10.665110 & 8.832716 & 1.215516 \\
\bottomrule
\end{tabular}
\end{table}

\end{document}